\documentclass[showpacs,preprintnumbers,amsmath,amssymb]{revtex4}
\usepackage{graphicx}
\usepackage{dcolumn}
\usepackage{bm}
\begin{document}
\title{A precisely feasible gauged model of chiral boson with its
BRST cohomological perspectives}
\author{Anisur Rahaman}
\email{anisur.associates@iucaa.ac.in, manisurn@gmail.com}
\affiliation{Durgapur Government College, Durgapur, PIN-713124
West Bengal, India}
\date{\today}
\begin{abstract}
We find that Siegel type chiral boson with a parameter-dependent
Lorentz non-covariant masslike term for the gauge fields to be
equivalent to the chiral Schwinger model with one parameter class
of Faddeevian anomaly if the model is described in terms of
Floreanini-Jackiw type chiral boson. By invoking the Wess-Zunino
field gauge-invariant reformulation is made. It has been shown
that the gauge-invariant model has the same physical content as
its gauge non-invariant ancestor had. The BRST invariant effective
action corresponding to this model has also been constructed. All
the nilpotent symmetries associated with the BRST symmetry along
with the bosonic, ghost, and discrete symmetries have been
systematically studied. We establish that the nilpotent charges
corresponding to these symmetries resemble the algebra of the de
Rham cohomological operators in differential geometry. In the
environment of conserved charges associated with the models, we
study the Hodge decomposition theorem on the compact manifold.
\end{abstract}
 \maketitle
\section{Introduction}
The chiral boson is the basic ingredient in the construction of
heterotic string theory \cite{ST1, ST2, ST3, ST4}. The theory of
quantum Hall effect additionally got important input from the
physics of chiral boson \cite{HAL1, HAL2}. The chiral boson
\cite{SIG, JSON, PPS} and the gauged model of it \cite{BEL} had
been advanced independently. It became determined that the gauged
model of the chiral boson and the chiral Schwinger had exciting
connections regardless of the structural differences between them
\cite{KH, PMS, SM2, ARC}. The Chiral Schwinger model is an
interacting model which includes fermion and gauge field and the
interaction among those is chiral in nature.  This model \cite{JR}
was extensively studied over the decades \cite{JR, RABIN, ROT1,
ROT2, PM, SM1} and attracted extra attention while it became
determined that it could be characterized through the expressions
of chiral boson \cite{KH, PMS, SM2, ARC}. The model can be
expressed in terms of chiral boson in view of the fact that in
$(1+1)$ dimension exact bosonization is possible. However, the
bosonization requires regularization which removes the hassles due
to non-unitarity for which it had to suffer for quite a long
period. Two independent regularizations are found within the
literature for this model. The initial one is called the
Jackiw-Rajaramn kind co-variant regularization \cite{JR} and the
latter one is Mitra kind non-covariant regularization \cite{PM}.
Mitra himself termed the anomaly in it as Faddeevian anomaly
\cite{FAD1, FAD2} in view of the fact that Gauss law commutator
for this version rendered a non-trivial contribution \cite{PM}.
Unlike Jackiw-Rajaraman's development, the masslike term for the
gauge field was independent of any parameter in Mitra's
development \cite{PM}. A few years later, a one-parameter involved
improvement of Faddeevian regularization became advanced in
\cite{WOT}. In this article, an attempt has been made to establish
the hidden link between the gauged chiral boson \cite{BEL} and the
bosonized version of chiral Schwinger with the one-parameter
involved Faddeevian anomaly described in \cite{WOT}. The
description of the model \cite{WOT} in terms of chiral boson was
pursued in our article \cite{ARC}. The investigation of the
symmetry belonging to this version would be instructive since it
is an anomalous gauge theory. The model itself has no gauge
symmetry, however, gauge symmetry can be restored by enlarging the
phasespace of the theory with the introduction of auxiliary
fields. These fields render their incredible service in restoring
the gauge symmetry keeping themselves allocated in the un-physical
sector of the theory.

Since it is an anomalous theory, the study of
Becchi-Rouet-Stora-Tyutin (BRST) is of interest since the BRST
formalism in the context of covariant canonical quantization of
gauge-invariant theories  \cite{BRS1, BRS2, BRS3, FIK, BAT1, BAT2,
BAT3, BAT4} play a pivotal role in the regime of formal field
theory. It ensures unitarity at any arbitrary order of
perturbation in the computations of physical processes. The gauge
theories endowed with the anti-BRST as well as anti-co-BRST
symmetries within the framework of BRST formalism can be shown to
provide a set of tractable physical examples for the Hodge theory
where the symmetries and the corresponding conserved charges
provide the physical realization of the de Rham cohomological
operators of differential geometry \cite{EGUCHI, HOLT, TBOM}.

Batalin Fradkin vilkovisky (BFV) \cite{BAT1, BAT2, BAT3, BAT4}
formalism and its applications in different field theoretical
models \cite{KIM1, KIM2, KIM3, SG, MIAO, ALEX1, ALEX2, AR1, ASN,
ASE, ARSS, AR3, ARS, BAR} has added a huge instructive and
illuminating information in the field theoretical regime. So the
attempt to construct the BRST invariant reformulations of this
gauged chiral boson having a non-covariant para meter involved
masslike term for gauge field taking the help of BFV formulation
will add a new and instructive contribution to the regime of
formal field theory.

The BRST cohomological aspects related to it will likewise be an
significant extension. So along with the nilpotent BRST symmetry,
we concentrate on the other nilpotent symmetries like ant-BRST,
co-BRST, and anti-co-BRST symmetry in this framework
systematically. There are a few investigations where endeavors
have been made for various models to show that the generators of
of the symmetries related to the BRST resembles the algebraic
structure of de Rham cohomological operator of differential
geometry \cite{RPM0, RPM1, RPM2, RPM3,RPM4, SUDHA, SUDHA1, ASN,
MALIK1, MALIK2, MALIK3, HODGE1, HODGE2, HODGE3, HODGE4, HODGE5}. A
unique endeavor in this manner is made here to analyze whether the
generators of these continuous symmetries satisfy algebra of de
Rham cohomological operators of differential geometry. The
foundation of the resemblance of the algebra of de Rham
cohomological operators have been tried to administrator with the
acquired insight from the past investigations. In the specific
circumstance of conserved charges associated with the models, we
study the Hodge decomposition theorem on the compact manifold and
found that the gauged chiral boson that contrasts with masslike
terms for the gauge field with a non-covariant parameter involved
mass-like terms have a place with the class of Hodge hypothesis

This article is organized as follows. Sec. II contains the
formulation of gauged Floreanini-Jackiw type chiral boson that
corresponds to the one-parameter class of Faddeevian anomaly. In
Sec. III, a review of the theoretical spectral of this model is
made. In Sec. IV, the gauge-invariant version is constructed with
the use of the Wess-Zumino field. In Sec. V, an equivalence
between and gauge non-invariant version is made. Sec. VI contains
a discussion of a similar type of model that contains one more
chiral degree of freedom in the theoretical spectrum. In Sec. VII,
like the previous cess with one less chiral degree of freedom here
also an equivalence is made between the gauge-invariant and the
gauge non-invariant version of this model. Sec VIII. is devoted to
the BFV quantization of gauged FJ type chiral boson with a
parameter-dependent non-covariant masslike term for gauge field.
Sec. IX, contains discussions over the extended BRST symmetries.
In Sec. X, the BRST Cohomological aspect of the theory is studied.
The final Sec. XI, contains a summary and discussions.

\section{Formulation of gauged Floreanini-Jackiw type chiral
boson that corresponds to one-parameter class of Faddeevian
anomaly} A gauged model of Siegel type chiral boson that
resemblesthe chiral Schwinger model with a one-parameter class of
Jackiw-Rajaraman type regularization was mentioned in \cite{ARC}.
An extension that follows naturally is that the gauge model of
Siegel type chiral boson with an appropriate selection of masslike
terms for the gauge field is equivalent to the gauged
Floreanini-Jackiw type chiral boson \cite{FJ} with a one-parameter
class of Faddeevian anomaly \cite{ARC}. To formulate that let us
proceed with the subsequent Lagrangian containing a suitable
parameter-dependent Lorentz non-covariant masslike term for the
gauge field.
\begin{eqnarray}
L_{CH}&=& \int dx[
\frac{1}{2}(\dot{\phi}^{2}-\phi'^{2})+e(\dot{\phi}+\phi')(A_{0}-A_{1})
+\frac{\tau}{2}[(\dot{\phi}
-\phi')+e(A_{0}-A_{1})]^{2}\nonumber \\
&+&\frac{1}{2}(\dot{A}-A'_{0})^{2}+\frac{1}{2}e^{2}(A^2_{0}+
2\alpha A_{1}A_{0} + (2\alpha-1)A^2_{1})]. \label{SIGCBB}
\end{eqnarray}
Here $\phi$ represents a scalar field. $A_0$and $A_1$ are the
twocomponents of the gauge field in $(1+1)$ dimension. Over-dot
and over-prime indicate the time and space derivatives
respectively. Note that there is a parameter-dependent Lorentz
non-covariant masslike term for the gauge field within Lagrangian
(\ref{SIGCBB}). Although the masslike term is lacking Lorentz
covariance, it ultimately renders an interesting and physically
sensible Lorentz invariant theory. What follows is the
illustration to ascertain the physical sensibility of the model.
To this endeavor, we need to work out the canonical momenta
corresponding to the fields $A_0$, $A_1$, $\phi$, and $\tau$
\begin{equation}
\frac{\partial L}{\partial\dot{A}_0}=\pi_0\approx 0, \label{MOM1}
\end{equation}
\begin{equation}
\frac{\partial
L}{\partial\dot{A}_1}=\pi_1=(\dot{A}_1-A'_0),\label{MOM2}
\end{equation}
\begin{equation}
\frac{\partial
L}{\partial\dot{\phi}}=\pi_{\phi}=(1+\tau)\dot{\phi}-\tau\phi'
+e(1+\tau)(A_{0}-A_{1}),\label{MOM3}
\end{equation}
\begin{equation}
\frac{\partial L}{\partial\dot{\tau}}=\pi_{\tau}\approx
0.\label{MOM4}
\end{equation}
The field $\tau$ is a lagrange multiplier field. Exploiting the
Legendre transformation $H= \pi_0\dot{A}_0 + \pi_1\dot{A}_1 +
\pi_{\phi}\dot{\phi} + \pi_{\tau}\dot{\tau} -L$ along with the use
of the expression of momenta (\ref{MOM1}), (\ref{MOM2}),
(\ref{MOM3}), (\ref{MOM4}) we compute the canonical Hamiltonian:
\begin{eqnarray}
H_{C}&=& \int dx {\cal H}_{C}= \int dx[
\frac{1}{2}\pi_{1}^{2}+\pi_{1}A'_{0}+\pi_{\phi}\phi'
-e(\pi_{\phi}+\phi')(A_{0}-A_{1})\nonumber \\
&+& e^2[(\alpha-1)A_{1}^{2}+ (\alpha+1)A_0A_1]+
\frac{1}{2(1+\tau)}(\pi_\phi-\phi')^2].
\end{eqnarray}
The equations (\ref{MOM1}) and (\ref{MOM4}) are the primary
constraint of the theory since there is no time derivative in
these equations. The preservation of these constraints leads to
some additional constraints or it fixes the velocity. Repeating
this preservation criterion on the usual constraint and the
forthcoming secondary constraints we find that the phasespace of
the system is endowed with the following five constraints.
\begin{equation}
{\cal C}_{1}=\pi_{0}\approx 0,
\end{equation}
\begin{equation}
{\cal C}_{2}=\pi_{\tau}\approx 0,
\end{equation}
\begin{equation}
{\cal C}_{3}=\pi'_{1}+e(\pi_{\phi}+\phi')+ (\alpha+1)A_1\approx 0,
\end{equation}
\begin{equation}
{\cal C}_{4}=\pi_{\phi}-\phi'\approx 0,
\end{equation}
\begin{equation}
{\cal C}_{5}=(\alpha+1)\pi_1+ 2\alpha(A'_0 +A'_1) \approx 0,
\end{equation}
We use a gauge fixing condition
\begin{equation}
{\cal C}_{6}=\tau+f(x)\approx 0.
\end{equation}
Note that  the constraint $\pi_{\tau}\approx 0$ is first class,
and it generates the Siegel gauge symmetry, by the
Anderson-Bergman algorithm, we can fix the gauge  ${\cal C}_{6} =
\tau+f(x)$ where $f(x)$ is an arbitrary function \cite{ARR}.
Therefore, we can formulate the generating functional
corresponding to the theory as follows:
\begin{eqnarray}
Z&=&\int |det[{\cal C}_k, {\cal C}_l]^{\frac{1}{2}}|
dA_{1}d\pi_{1} d\phi d\pi_{\phi} d\Lambda
d\pi_{\Lambda}dA_{0}d\pi_{0}e^{i\int
d^2x(\pi_{1}\dot{A_{1}}+\pi_{\phi}\dot{\phi}+\pi_{\tau}\dot{\tau}
+\pi_{0}\dot{A_{0}}-{\cal H}_{C})}\nonumber\\
&&\times \delta({\cal C}_{1})\delta({\cal C}_{2})\delta({\cal
C}_{3})\delta({\cal C}_{4})\delta({\cal C}_{5})\delta({\cal
C}_{6}).
\end{eqnarray}
The subscripts $k$ and $l$ run from $1$ to $6$. The simplification
by the use of Gaussian integral leads us to
\begin{equation}
Z=\int d\phi dA_{1} e^{i\int d^2x {\cal L}_{CH}},
\end{equation}
where
\begin{equation}
{\cal L}_{CH}
=\dot{\phi}\phi'-\phi'^{2}+2e\phi'(A_{0}-A_{1})-e^2[(\alpha-1)A_{1}^{2}-
(\alpha+1)A_0A_1] +\frac{1}{2}(\dot{A_{1}}-A'_{0})^{2}.
\label{LCH0} \end{equation}
Thus, it manifests transparently that
the Lagrangian (\ref{LCH0}) is the appropriate gauged Lagrangian of
Siegel type chiral boson that corresponds to the gauged model of
chiral boson with the one-parameter class of Floreanini-Jackiw
type gauged chiral boson which can be generated from the chiral Schwinger
model with one-parameter class Faddeevian anomaly \cite{WOT}
introducing a chiral constraint in the phase space of the theory
\cite{ARC}.
\section{Review of the theoretical spectrum}
In this section, we describe the theoretical spectrum of this
system in brief. The Lagrangian density with which we begin our
analysis to find out the phasespace structure of the theory is
\begin{equation}
{\cal L}_{CH}
=\dot{\phi}\phi'-\phi'^{2}+2e\phi'(A_{0}-A_{1})-e^2[(\alpha-1)A_{1}^{2}-
(\alpha+1)A_0A_1] +\frac{1}{2}(\dot{A_{1}}-A'_{0})^{2}.
\label{LCH00}
\end{equation}
From the standard definition, the momenta  $\pi_\phi$, $\pi_0$,
and $\pi_1$ corresponding to the fields $\phi$, $A_0$, and $A_1$
are obtained:
\begin{equation}
\pi_\phi = \phi',\label{MO1}
\end{equation}
\begin{equation}
\pi_0 = 0,\label{MO2}
\end{equation}
\begin{equation}
\pi_1 = \dot A_1 - A_0'.\label{MO3}
\end{equation}
Using the above equations (\ref{MO1}), (\ref{MO2}), and
(\ref{MO3}), it is straightforward to obtain the canonical
Hamiltonian which reads
\begin{equation}
H_C = \int dx[\frac{1}{2}\pi_1^2 + \pi_1A_0' + \phi'^2 - 2e(A_0 -
A_1)\phi' + \frac{1}{2}e^2[2(\alpha -1)A_1^2 + 2(\alpha +
1)A_0A_1)].
\end{equation}
Equation (\ref{MO1}) and (\ref{MO2}) are the primary constraints
of the theory. Therefore, the effective Hamiltonian is given by
\begin{equation}
H_{EFF} = H_C + {\it u}\pi_0 + {\it v}(\pi_\phi - \phi'),
\end{equation}
where ${\it u}$ and ${\it v}$ are two arbitrary Lagrange
multipliers. The constraints obtained in (\ref{MO1}) and
(\ref{MO2}) have to be preserved in time to have a consistent
physical theory. The preservation of the constraint (\ref{MO2}),
gives the Gauss law of the theory:
\begin{equation}
{\cal G} = \pi_1' + 2e\phi' +  e^2(1 + \alpha)A_1 = 0.
\label{GAUS}
\end{equation}
The consistency  criterion of the constraint (\ref{MO1}) although
renders no new  constraint  it determines the velocity ${\it v}$
which is given by
\begin{equation}
{\it v} = \phi' - e(A_0 - A_1). \label{VEL}
\end{equation}
The  Gauss law constraint, entails $\dot{\cal G} = 0$, to get
preserved in time that results in a new constraint
\begin{equation}
(1 + \alpha)\pi_1 + 2\alpha (A_0' +A_1') = 0.\label{FINC}
\end{equation}
The preservation of the constraint (\ref{FINC}) does not give any
new constraint. So we find that the phasespace of the theory is
embedded with the constraints (\ref{MO1}), (\ref{MO2}),
(\ref{GAUS}), and (\ref{FINC}) and are all of these are weak
conditions unto this stage. If the constraints are treated as
strong conditions the following reduced Hamiltonian results in.
\begin{equation}
H_R =  \int dx[\frac{1}{2}\pi_1^2 + \frac{1}{4e_2} \pi_1'^2 +
\frac{1}{2}(\alpha - 1) \pi_1'A_1 + \frac{1}{2}e^2[(1 + \alpha)^2
- 4\alpha)]A_1^2]. \label{RHAM}
\end{equation}
Since the constraints are treated here as strong condition to
obtain the reduced Hamiltonian the usual Poisson's bracket becomes
inadequate however, there is a remedy: the reduced Hamiltonian is
known to be consistent with the Dirac brackets \cite{DIR}. The
Dirac brackets between the fields describing the reduced
Hamiltonian $H_R$ are found out to be
\begin{equation}
[A_1(x), A_1(y)]^* =\frac {1}{2e^2}\delta'(x-y), \label{DR1}
\end{equation}
\begin{equation}
[A_1(x), \pi_1(y)]^* = \frac{(\alpha -1)}
{2\alpha}\delta(x-y),\label{DR2}
\end{equation}
\begin{equation}
[\pi_1(x), \pi_1(y)]^* = -\frac{(1+\alpha)^2}
{4\alpha}e^2\epsilon(x-y).\label{DR3}
\end{equation}
Using the reduced Hamiltonian (\ref{RHAM}), and the Dirac brackets
(\ref{DR1}), (\ref{DR2}), and (\ref{DR3}), a little algebra leads
us to obtain the first-order equations of motion for $A_1$ and
$\pi_1$:
\begin{equation}
\partial_{-}\pi_1 = \frac{(\alpha-1)^2 }{\alpha} e^2A_1, \label{REQ1}
\end{equation}
\begin{equation}
\partial_{+}A_1 = \frac{(\alpha-1)} {2\alpha} \pi_1
+ \frac{1}{2\alpha}(\alpha+ 1)A_1',\label{REQ2}
\end{equation}
and  these first-order equations of motion reduce to the following
second-order equation after little simplification. It is
straightforward to see that the above two equations (\ref{REQ1})
and (\ref{REQ2}) satisfy a Klein-Gordon type Equation
\begin{equation}
(\Box - \frac{(\alpha-1)}{\alpha})\pi_1 = 0.\label{SPEC}
\end{equation}
The equation (\ref{SPEC}), represents a massive boson with square
of the mass $m^2 = -e^2\frac{(1-\alpha)^2}{\alpha }$. It is
evident that the parameter $\alpha$ must be negative for the mass
of the boson to be physical. Unlike the Abreu {\it et al.},
\cite{WOT} there is no massless degree of freedom in this
situation since the phasespace of this theory contains one more
constraint. After this brief review of the theoretical spectrum
let us proceed to study the gauge symmetric properties of this
model.
\section{Gauge invariant version with the Wess-Zumino}
The model in its usual phasecpace is not gauge-invariant. The use
of the Wess-Zumino field helps to get a gauge-invariant Lagrangian
in the extended phasespace. The gauge-invariant Lagrangian
corresponding to this gauged chiral boson that resembles the
chiral Schwinger model with a one-parameter class Faddeevian
anomaly is given by
\begin{equation}
L_{CHI}=\int dx [{\cal L}_{CH}+{\cal L}_{WZ}],
\end{equation}
where ${\cal L}_{WZ}$ refers to the Wess-Zumino term \cite{WESS}
\begin{equation}
{\cal L}_{WZ} = \alpha(\dot{\omega}\omega' + \omega'^2) +
e(\alpha+1)(A_0\omega'-A_1\dot{\omega}) -2e\alpha (A_0+A_1)\omega.
\end{equation}
Therefore, the total Lagrangian density reads
\begin{eqnarray}
{\cal L}_{CHI}
&=&\dot{\phi}\phi'-\phi'^{2}+2e\phi'(A_{0}-A_{1})-e^2[(\alpha-1)A_{1}^{2}+
(\alpha+1)A_0A_1] +\frac{1}{2}(\dot{A_{1}}-A'_{0})^{2}
\nonumber\\
&+&  \alpha(\dot{\omega}\omega' + \omega'^2) +
e(\alpha+1)(A_0\omega'-A_1\dot{\omega}) -2e\alpha
(A_0+A_1)\omega'.\label{FJCBI}
\end{eqnarray}
Here $\omega$ represents the Wess-Zumino field \cite{WESS}. The
momenta corresponding to the field $A_0$, $A_1$, $\phi$, and
$\omega$ respectively are
\begin{equation}
\frac{\partial L_{CHI}}{\partial\dot{A}_0}=\pi_0\approx 0,
\label{IMOM1}
\end{equation}
\begin{equation}
\frac{\partial
L_{CHI}}{\partial\dot{A}_1}=\pi_1=(\dot{A}_1-A'_0),\label{IMOM2}
\end{equation}
\begin{equation}
\frac{\partial L_{CHI}}{\partial\dot{\phi}}=\phi' ,\label{IMOM3}
\end{equation}
\begin{equation}
\frac{\partial L_{CHI}}{\partial\dot{\omega}}=\pi_{\omega} =
\alpha \omega' - e(1+\alpha)A_1 \approx 0.\label{IMOM4}
\end{equation}
The canonical Hamiltonian is obtained using the equation
(\ref{IMOM1}), (\ref{IMOM2}), (\ref{IMOM3}), and (\ref{IMOM4}) by
the use of a Legendre transformation:
\begin{equation}
H_{C}=\int dx[
\pi_{1}\dot{A_{1}}+\pi_{0}\dot{A_{0}}+\pi_{\phi}\dot{\phi}
+\pi_{\omega}\dot{\omega}]-L_{CHI}]. \label{CHAM4}
\end{equation}
Therefore, the effective Hamiltonian reads
\begin{eqnarray}
H_{CHI}&=& \int dx[\frac{1}{2}\pi_{1}^{2}+\pi_{1} A'_{0}
+\phi'^2-2e(A_{0}-A_{1})\phi'- e^{2}[(\alpha-1) A_{1}^{2}+ (\alpha
+1)A_0A_1] \nonumber \\
&-& \alpha\omega'^2 -e(1+\alpha)A_0\omega'
+2e\alpha(A_0+A_1)\omega' + w_1\pi_0 + w_2(\pi_\phi - \phi') + w_3
[\pi_\omega -\alpha\omega' + e(1+\alpha)A_1]. \label{EHAM4}
\end{eqnarray}
The Gauss law constraint of the theory is computed using the
preservation of the constraint (\ref{IMOM1}), that comes out to be
\begin{equation}
G =\pi'_{1}+ 2e\phi' + e(1-\alpha)\omega' +
e^2(\alpha+1)A_1\approx 0.
\end{equation}
Here $w_1$, $w_2$, and $w_2$ are  the Lagrange multipliers having
dimension of velocity The velocities $w_2$ and $w_2$  are found
out to be
\begin{equation}
w_2= \phi' -e(A_0 -A_1),
\end{equation}
\begin{equation}
w_3=-\omega' + e(A_0 + A_1).
\end{equation}
The velocity $w_1$ however remains undetermined. The preservation
of the Gauss law constraint leads to a new constraint
\begin{equation}
\dot{G}= (\alpha+1)\pi_1 + 2\alpha(A_0+A_1)' \approx0.
\end{equation}
So, it appears that the gauge-invariant system has the following
five constraints:
\begin{equation}
{\cal K}_{1}=\pi_{0}\approx 0,
\end{equation}
\begin{equation}
{\cal K}_{2}=\pi_{\phi}-\phi'\approx 0,
\end{equation}
\begin{equation}
{\cal K}_{3}=\pi_{\omega}-\alpha \omega' + e(1+\alpha)A_1 \approx
0,
\end{equation}
\begin{equation}
{\cal K}_{4}=\pi'_{1}+ 2e\phi' + e(1-\alpha)\omega' +
e^2(\alpha+1)A_1\approx 0,
\end{equation}
\begin{equation}
{\cal K}_{5}=(\alpha+1)\pi_1+ 2\alpha(A'_0 +A'_1) \approx 0.
\end{equation}
Our next task is to make an equivalence between the
gauge-invariant and gauge non-invariant version of the theory
since gauge-invariant is made here in the extended phasespace with
the introduction of a Wess-Zumino field.
\section{To make an equivalence between the gauge invariant and
gauge non-invariant version} To make  an equivalence between the
gauge invariant and the gauge non-invariant interpretation
corresponding to this model let us proceed with the gauge
symmetric Lagrangian. So we add up the Wess-Zumino term
\cite{WESS} with the usual Lagrangian.
\begin{equation}
L_{CHI}=\int dx [{\cal L}_{CH}+{\cal L}_{WZ}]
\end{equation}
\begin{eqnarray}
{\cal L}_{CHI}
&=&\dot{\phi}\phi'-\phi'^{2}+2e\phi'(A_{0}-A_{1})-e^2[(\alpha-1)A_{1}^{2}+
(\alpha+1)A_0A_1] +\frac{1}{2}(\dot{A_{1}}-A'_{0})^{2}
\nonumber\\
&+&  \alpha(\dot{\omega}\omega' + \omega'^2) +
e(\alpha+1)(A_0\omega'-A_1\dot{\omega}) -2e\alpha
(A_0+A_1)\omega'\label{FJCBIP}
\end{eqnarray}
The canonical Hamiltonian in this situation reads
\begin{eqnarray}
H_{CHI}&=& \int dx[\frac{1}{2}\pi_{1}^{2}+\pi_{1} A'_{0}
+\phi'^2-2e(A_{0}-A_{1})\phi'- e^{2}[(\alpha-1) A_{1}^{2}+ (\alpha
+1)A_0A_1] \nonumber \\
&-& \alpha\omega'^2 -e(1+\alpha)A_0\omega'
+2e\alpha(A_0+A_1)\omega']
\end{eqnarray}
We will now follow the formalism developed in the article
\cite{FALCK} to establish the required equivalence.  To ensure it,
we require two gauge fixing conditions. The appropriate gauge
fixing conditions are
\begin{equation}
{\cal K}_{6}=\omega' \approx 0,
\end{equation}
\begin{equation}
{\cal K}_{7}=\pi_\omega+ e(\alpha+1)A_1\approx 0.
\end{equation}
There were  five constraints in the phasespace of the theory.
Those five constraints along with these two gauge fixing
conditions form a second class set. It enables us to write down
the generating functional:
\begin{eqnarray}
Z&=& \int[det[{\cal K}_{m},~{\cal
K}_{n}]]^{\frac{1}{2}}dA_{1}d\pi_{1} d\phi d\pi_{\phi}
dA_{0}d\pi_{0} d\omega d\pi_{\omega}e^{i\int d^2x
(\pi_{1}\dot{A_{1}}+\pi_{\phi}\dot{\phi}+\pi_{\omega}\dot{\omega}
+\pi_{0}\dot{A_{0}}-H_{C})}
\nonumber\\
&& \times \delta({\cal K}_{1})\delta({\cal K}_{2})\delta({\cal
K}_{3})\delta({\cal K}_{4}\delta({\cal K}_{5})\delta({\cal K}_{5})
\delta({\cal K}_{7}). \label{GENF}
\end{eqnarray}
Here $m$ and $n$ both runs from $1$ to $7$. Integrating out of the
fields $\omega$ and $\pi_{\omega}$ we find that equation
(\ref{GENF}) reduces to
\begin{eqnarray}
Z&=& N\int dA_{1}d\pi_{1} d\phi d\pi_{\phi}dA_{0}d\pi_{0}e^{i\int
d^2x(\pi_{1}\dot{A_{1}}+\pi_{\phi}\dot{\phi}
+\pi_{0}\dot{A_{0}}-{\cal H}_{GSF})}\nonumber\\
&&\times \delta(\tilde{{\cal K}}_{1})\delta(\tilde{{\cal
K}}_{2})\delta(\tilde{{\cal K}}_{4})\delta(\tilde{{\cal K}}_{5}).
\end{eqnarray}
where
\begin{equation}
\tilde{{\cal K}}_{1} = F_{1}=\pi_{0}\approx 0,
\end{equation}
\begin{equation}
\tilde{{\cal K}}_{2}= F_{2}=\pi_{\phi}-\phi'\approx 0,
\end{equation}
\begin{equation}
\tilde{{\cal K}}_{4}= \pi'_{1}+ 2e\phi'+ e^2(\alpha+1)A_1\approx
0,
\end{equation}
\begin{equation}
\tilde{{\cal K}}_{5}= \tilde{{\cal K}}_{f}=(\alpha+1)\pi_1+
2\alpha(A'_0 +A'_1)\approx 0.
\end{equation}
These are the usual set  of constraints of the gauge non-invariant
version of the theory and the corresponding Hamiltonian is
\begin{equation}
{\cal H}_{GSF} = \frac{1}{2}\pi_{1}^{2}+\pi_{1} A'_{0}
+\phi'^2-2e(A_{0}-A_{1})\phi'- e^{2}[(\alpha-1) A_{1}^{2}+ (\alpha
+1)A_0A_1.
\end{equation}
Again integrating out of the momenta $\pi_0$, $\pi_1$, and
$\pi_\phi$ we land onto
\begin{equation}
Z=\tilde{N}\int d\phi dA_{1} e^{i\int d^2x {\cal L}_{GSF}},
\end{equation}
where
\begin{equation}
 {\cal L}_{GSF}=
\dot{\phi}\phi'-\phi'^{2}+2e\phi'(A_{0}-A_{1})-e^2[(\alpha-1)A_{1}^{2}-
(\alpha+1)A_0A_1] +\frac{1}{2}(\dot{A_{1}}-A'_{0})^{2}.
\label{FJCB}
\end{equation}
Note that the system now contains the usual five constraints
$\tilde{{\cal K}} _{1}$, $\tilde{{\cal K}} _{2}$, $\tilde{{\cal
K}}_{4}$, and $\tilde{{\cal K}} _{5}$ and the Lagrangian density
${\cal }L_{GSF}$ is identical to the usual Lagrangian $L_{CH}$
having the same Hamiltonian $H_{GSF}= H_{R}$. So the gauge
invariant Lagrangian maps onto the gauge non-invariant Lagrangian
described in the usual phasespace. It also ensures that the
physical contents in both gauge-invariant and gauge non-invariant
versions are identical.
\section{A theory that contains one more chiral degrees of freedom}
An alternative description of this theory is possible where the
theory contains an extra chiral degree of freedom. If a chiral
constraint is imposed in it by hand it transforms into the theory
described in the earlier Sections \cite{ARC}. The theoretical
spectra corresponding to the  theory having  one more chiral
degree of freedom  was discussed in \cite{WOT}.
\begin{eqnarray}
L_{CB}&=& \int dx[
\frac{1}{2}(\dot{\phi}^{2}-\phi'^{2})+e(\dot{\phi}+\phi')(A_{0}-A_{1})
+\frac{1}{2}(\dot{A}-A'_{0})^{2}+\frac{1}{2}e^{2}(A^2_{0}+ 2\alpha
A_{1}A_{0} + (2\alpha-1)A^2_{1}].
\end{eqnarray}
The following is a brief review of the phasespace structure of
this theory. The momenta of the fields describing  the Lagrangian
are given by
\begin{equation}
\pi_0 = 0,\label{MOMN1}
\end{equation}
\begin{equation}
\pi_1 = \dot A_1 - A_0'.\label{MOMN2}
\end{equation}
\begin{equation}
\pi_\phi = \dot{\phi} -e(A_0-A_1),\label{MOMN3}
\end{equation}
It has been found that the theory contains three second class
constraints.
\begin{equation}
\pi_0 \approx 0,\label{CON1}
\end{equation}
\begin{equation}
\pi_1' +e(\pi_\phi +\phi')+ e^2(1+\alpha)A_1\approx 0,\label{CON2}
\end{equation}
\begin{equation}
(1+\alpha)\pi_1 + 2e\alpha(A_0+A_1)'\approx 0. \label{CON3}
\end{equation}
The theoretical spectrum is given by the flowing to second order
equations of motion \cite{WOT}
\begin{equation}
(\Box-e^2\frac{(\alpha-1)^2}{\alpha})\pi_1  = 0,
\end{equation}
\begin{equation}
\dot{\cal F}+{\cal F}' = 0,
\end{equation}
 where
${\cal F} = \phi -\frac{1}{e}\frac{\alpha}{1+\alpha}(\dot{A}_1 +
A_1')$. It indicates that theoretical spectra of the theory
contain one massive boson along with a massless chiral degree of
freedom.
\section{Gauge invariant version and making an equivalence with gauge variant version}
Like the previous case introducing the Wess-Zumino field, a
gauge-invariant version is possible to construct as follows. The
gauge-invariant version in the extended phasespace is given by
\begin{eqnarray}
L_{CBI}&=& \int
dx[\frac{1}{2}(\dot{\phi}^{2}-\phi'^{2})+e(\dot{\phi}+\phi')(A_{0}-A_{1})
+\frac{1}{2}(\dot{A}-A'_{0})^{2}+\frac{1}{2}e^{2}(A^2_{0}+ 2\alpha
A_{1}A_{0} + (2\alpha-1)A^2_{1}]
\nonumber \\
&+&\alpha(\dot{\omega}\omega'+\omega'^{2})+2e\alpha(A_{0}+A_{1})\omega'
+(\alpha+1)(A_1\dot{\omega}- A_0\omega'). \label{CBA}
\end{eqnarray}
The terms containing the field $\omega$ in Eqn. (\ref{CBA}) is
known as Wess-Zumino term and $\omega$ is the corresponding
Wess-Zumino field \cite{WESS}. To show the equivalence between
these two model we need to calculate the momenta corresponding to
the field $A_0$, $A_1$, $\phi$, and $\omega$ like the previous
case.
\begin{equation}
\frac{\partial L_{CBI}}{\partial\dot{A}_0}=\pi_0\approx 0,
\label{AMOM1}
\end{equation}
\begin{equation}
\frac{\partial
L_{CBI}}{\partial\dot{A}_1}=\pi_1=(\dot{A}_1-A'_0),\label{AMOM2}
\end{equation}
\begin{equation}
\frac{\partial L_{CBI}}{\partial\dot{\phi}}=\pi_\phi + e(A_0-A_1),
\label{AMOM3}
\end{equation}
\begin{equation}
\frac{\partial L_{CBI}}{\partial\dot{\omega}}=\pi_{\omega} =
\alpha \omega' - e(1+\alpha)A_1 \approx 0.\label{AMOM4}
\end{equation}
Here $\pi_\phi$, $\pi_0$, and $\pi_1$, and $\pi_\omega$ are the
momentum corresponding to the field $A_0$, $A_1$, $\phi$, and
$\omega$ The canonical Hamiltonian is is obtained through a
Legendre transformation along with using the definition of
momenta:
\begin{eqnarray}
H_{CBI}&=& \int dx[\frac{1}{2}(\pi_{1}^{2}+ +\phi'^2 + \pi_\phi^2)
+ \pi_{1} A'_{0} - e(A_{0}-A_{1})(\pi_\phi +\phi') -
e^{2}[(\alpha-1) A_{1}^{2}+ (\alpha
+1)A_0A_1] \nonumber \\
&-& \alpha\omega'^2 -e(1+\alpha)A_0\omega'
+2e\alpha(A_0+A_1)\omega'].
\end{eqnarray}
Note that $\pi_0\approx 0$ and $\pi_{\omega} - \alpha \omega' +
e(1+\alpha)A_1 \approx 0$ are the primary constraints of the
theory. The system contains two more constraints. The constraints
are explicitly given by
\begin{equation}
C_{1}=\pi_{0}\approx 0,
\end{equation}
\begin{equation}
 C_{2}=\pi_{\omega} - \alpha \omega' +
e(1+\alpha)A_1, \approx 0,
\end{equation}
\begin{equation}
C_{3}= \pi'_{1}+ e(\pi_\phi + \phi')+ e^2(\alpha+1)A_1 +
(1-\alpha)\omega'.
\end{equation}
It is convenient to write down the generating functional to make
an equivalence keeping the required variables only, integrating
out the rest of them. Here we need some gage fixing conditions as
suggested in the article \cite{FALCK}. The gauge fixing conditions
that suit here are
\begin{equation}
C_4=\omega' \approx 0
\end{equation}
\begin{equation}
C_5=\pi_\omega -( 1+\alpha)\pi_1 +2\alpha(A_0 + A_1)'+ e(1+
\alpha)A_1 \approx 0
\end{equation}
We are now in a position to formulate the generating functional of
the theory. It reads
\begin{eqnarray}
Z&=&\int |det[C_k, C_l]^{\frac{1}{2}}| dA_{1}d\pi_{1} d\phi
d\pi_{\phi} d\omega d\pi_{\omega}dA_{0}d\pi_{0}e^{i\int
d^2x(\pi_{1}\dot{A_{1}}+\pi_{\phi}\dot{\phi}+\pi_{\Lambda}\dot{\Lambda}
+\pi_{0}\dot{A_{0}}-{\cal H}_{C})}\nonumber\\
&&\times
\delta(C_{1})\delta(C_{2})\delta(C_{3})\delta(C_{4})\delta(C_{5}).
\end{eqnarray}
The subscripts $k$ and $l$ runs from $1$ to $5$. After
simplification by the use of gaussian integral we land on to where
the Liouville measure $[{\cal D}\mu] = d\pi_\phi d\phi d\pi_1dA_1
d\pi_0 dA_0 d\pi_\omega d\omega$, and $m$ and$n$ run from $1$ to
$5$. After integrating out of the field $\omega$ and
$\pi_{\omega}$, we find that the generating functional turns into
\begin{equation}
Z= N \int[d\bar\mu]e^{i\int d^2x[\pi_1\dot A_1 + \pi_0 \dot A_0
+\pi_\phi \dot\phi+  - \tilde H_{CBS}]} \delta(\tilde{C}_1)
\delta(\tilde{C}_2)\delta(\tilde{C}_3), \label{CCZ}
\end{equation}
where $[{\cal D}\tilde\mu] = d\pi_\phi d\phi d\pi_1dA_1 d\pi_0
dA_0 $, and $N$ is a normalization constant having no significant
physical importance, and $\tilde H_{CNS}$ is given by
\begin{eqnarray}
\tilde H_{CBS}=\frac{1}{2}(\pi_1^2+\phi'^2+
\pi_{\phi}^2)+\pi_1A_0' +
e(\pi_\phi+\phi')(A_0-A_1)+e^2[(\alpha-1)A_{1}^{2}+
(\alpha+1)A_0A_1],\label{BARH}
\end{eqnarray}
and
\begin{equation}
\tilde{C}_1=\pi_0,
\end{equation}
\begin{equation}
\tilde{C}_2= \pi_1' +e(\pi_\phi +\phi')+ e^2(1+\alpha)A_1,
\end{equation}
\begin{equation}
\tilde{C}_3= (1+\alpha)\pi_1' + 2\alpha(A_0+A_1),
\end{equation}
which are the constraints of the theory in the gauge non-invariant
situation as given in (\ref{CON1}), (\ref{CON2}), and
(\ref{CON3}). We land onto the required result after integrating
out of the momenta $\pi_\phi$, $\pi_1$ and $\pi_0$:
\begin{equation}
Z=N\int d\phi dA_{1}DA_{0} e^{i\int d^2x {\cal L}_{CB}}
\end{equation}
where
\begin{equation}
{\cal L}_{CB}
=\frac{1}{2}(\dot{\phi}^{2}-\phi'^{2})+e(\dot{\phi}+\phi')(A_{0}-A_{1})-e^2[(\alpha-1)A_{1}^{2}-
(\alpha+1)A_0A_1] +\frac{1}{2}(\dot{A_{1}}-A'_{0})^{2}
\label{FJCBB}
\end{equation}
So it is now transparent that the Lagrangian (\ref{FJCBB}) is the
appropriate gauge-invariant Lagrangian density corresponding the
Lagrangian of bosonized chiral Schwinger model with one-parameter
class of Faddeevian anomaly \cite{WOT}.
\section{BRST formulation of gauged FJ type chiral
boson with non-covariant masslike term for gauge field} BRST
formulation is instructive for any theory since it ensures the
unitarity and renormalization of a physical theory. Batalin,
Fradkin, and vilkovisky (BFV) formalism serves as an important
tool to construct BRST invariant reformulation. We have got here a
scope to exploit BFV formalism to write down the BRST invariant
reformulation of the two nearly similar models, but having the
differences in the number of constraints they possessed in the
phasespces. What follows next is an attempt to write down the BRST
invariant Lagrangian for the model described in equation
(\ref{FJCB}) using the flavors of BFV formulation as it was found
in the article Fujiwara Igarashi and Kubo \cite{FIK}. It was shown
that using the improved version of BFV formalism the BF field
needed to make a theory first-class turns into Wess-Zumino scalar
\cite{FIK}. So what we need is the first-class theory
corresponding to this Lagrangian introducing the appropriate
Wess-Zumino term. By the extension of phasespace by Wess-Zumino
field $\omega$ we get the following gauge-invariant structure
corresponding to this model:
\begin{equation}
{\cal L}_{WZ} = \alpha(\dot{\omega}\omega' + \omega'^2) +
e(\alpha+1)(A_0\omega'-A_1\dot{\omega}) -2e\alpha (A_0+A_1)\omega
\end{equation}
Therefore, the total Lagrangian reads
\begin{eqnarray}
 {\cal L}_{CHI}
&=&\dot{\phi}\phi'-\phi'^{2}+2e\phi'(A_{0}-A_{1})-e^2[(\alpha-1)A_{1}^{2}+
(\alpha+1)A_0A_1] +\frac{1}{2}(\dot{A_{1}}-A'_{0})^{2}
\nonumber\\
&+&  \alpha(\dot{\omega}\omega' + \omega'^2) +
e(\alpha+1)(A_0\omega'-A_1\dot{\omega}) -2e\alpha
(A_0+A_1)\omega'\label{FJCBI1}
\end{eqnarray}
The Lagrangian density  has already been given in equation
(\ref{FJCBI}). The momenta corresponding to the field $A_0$,
$A_1$, $\phi$ and $\sigma$ respectively are calculated in
equations (\ref{IMOM1}), (\ref{IMOM2}),(\ref{IMOM3}), and
(\ref{IMOM4}). The canonical and the effective Hamiltonian are
computed in equations (\ref{CHAM4}) and  (\ref{EHAM4})
respectively. The Gauss law constraint of the theory reads
\begin{equation}
{\tilde G} =\pi'_{1}+ 2e\phi' + e(1-\alpha)\omega' +
e^2(\alpha+1)A_1\approx 0.
\end{equation}
In Sec. IV, we have seen that  the gauge invariant system has the
five constraints. Let us now follow the result of the article
\cite{FIK} to write down the BRST invariant effective action
without going to the formal construction using BFV formalism. With
the information of the article \cite{FIK} we can immediately write
down a BRST invariant effective action since BF fields turns in to
Wess-Zumino with appropriate choice of gauge fixing:
\begin{eqnarray}
 S_{eff}&=&\int
d^2x[\pi_{\phi}\dot{\phi}+\pi_{1}\dot A_{1} +\pi_{0}\dot
A_{0}+\pi_{\omega}\dot {\omega}-[\frac{1}{2}\pi_{1}^{2}+\pi_{1}
A'_{0} +\phi'^2-2e(A_{0}-A_{1})\phi'- e^{2}[(\alpha-1) A_{1}^{2}+
(\alpha
+1)A_0A_1] \nonumber \\
&-& \alpha\omega'^2-e(1+\alpha)A_0\omega'
+2e\alpha(A_0+A_1)\omega']+\dot{{\cal C}}{\cal P}+{\dot{\bar {\cal
C}}}{\bar P} -[{\cal Q}_b, \Psi]]. \label{FJCA}
\end{eqnarray}
Here ${\tilde Q}_b$ and  $\Psi$ stand for the BRST charge and
fermionic gauge fixing term respectively. The fields ${\cal C}$
and ${\cal P}$ is a pair of canonically conjugate ghost fields
with ghost numbers $1$ and $-1$ respectively, and the fields
${\bar {\cal C}}, {\bar {\cal P}}$ is a pair of canonically
conjugate anti-ghost fields with ghost numbers $-1$ and $1$
respectively.  The constraints ${\cal K} _f= {\cal K}_4 -e{\cal
K}_3 + e{\cal K}_2$ and ${\cal K}_1$ form a first-class set and
these two form the generator of the gauge transformation. The
gauge transformations corresponding to the fields $A_\mu, \phi$,
and $\omega$ are the following:
\begin{equation}
A_\mu\to A_\mu + \frac{1}{e}\partial_{\mu} \lambda,~~~\phi\to
\phi+\lambda, ~~~ \omega\to\omega-\lambda,
\end{equation}
which keep the Lagrangian invariant.  The constraints ${\cal K}_f=
{\cal K}_4 -e{\cal K}_3 + e{\cal K}_2$ and ${\cal K}_1$ constitute
the BRST charge ${\tilde Q}_b$ for this theory:
\begin{equation}
{\tilde Q}_b=\frac{i}{e}{\cal
C}[\pi_{1}^{\prime}+e(\pi_{\phi}-\pi_{\omega})+e(\phi^{\prime}+\omega^{\prime})]
-\frac{i}{e}\pi_{0}\bar {\cal P},
\end{equation}
The fermionic gauge fixing term in this situation is
\begin{equation}
\Psi=e[{\cal P} A_0+\bar{{\cal
C}}(\frac{1}{2}\pi_0+\partial_1A^1)].
\end{equation}
The BRST transformations of the fields constituting the effective
action $S_{eff}$ get generated from the BRST charger $Q_b$, which
are given by
\begin{eqnarray}
s_b\phi=-{\cal C}, s_b\omega={\cal C},  s_b\pi_\phi=-{\cal
C}^\prime, s_b\pi_\omega=-{\cal C}^\prime,
s_bA_0=\frac{1}{e}\dot{{\cal C}}, s_bA_1=\frac{1}{e}{\cal C}^\prime, \nonumber \\
s_b{\cal C}=0,
s_bp=\frac{1}{e}[\pi_{1}^{\prime}+e(\pi_{\phi}-\pi_{\omega})+e(\phi^{\prime}+\omega^{\prime})],
s_b\bar{{\cal C}}=\frac{1}{e}\pi_0, s_b\bar{{\cal P}}=0.
\label{BRT}
\end{eqnarray}
With the use of equation (\ref{BRT}) we find that
\begin{eqnarray}
[Q_b,\Psi] &=&
A_0[\pi_{1}^{\prime}+e(\pi_{\phi}-\pi_{\omega})+e(\phi^{\prime}+\omega^{\prime})]
+\pi_0(A_1^\prime-\frac{1}{2}\pi_0)+\bar{\cal P}{\cal P
}+\bar{\cal C}^{\prime}{\cal C}^{\prime}\nonumber\\
&\approx & -\pi_0\partial_1 A^1-\frac{1}{2}\pi_0^{2}+\bar{\cal
P}{\cal P}+\bar{\cal C}^{\prime}{\cal C}^{\prime}.
\end{eqnarray}
So, the generating functional corresponding to the BRST invariant
effective action takes the form
\begin{eqnarray}
Z&=&\int D\phi e^{S_{eff}},
\end{eqnarray}
where
\begin{eqnarray}
S_{eff}&=&\int d^2x[\pi_{\phi}\dot{\phi}+\pi_{\omega}\dot
{\omega}+\pi_{0}(\dot A_{0}-A_{1}^{\prime})
+\frac{1}{2}\pi_{0}^{2}-\frac{1}{2}\pi_{1}^{2}+\pi_{1}(\dot A_{1}-A_{0}^{\prime})\nonumber\\
&-&\phi^{\prime2}+2e\phi^{\prime}(A_{0}-A_{1}) - e^{2}[(\alpha-1)
A_{1}^{2}+ (\alpha
+1)A_0A_1]-\omega^{\prime2}+2e\omega^{\prime}(A_{0}+A_{1})
\nonumber\\
&+&\dot{{\cal C}}{\cal P} +\dot{\bar {\cal C}}\bar{{\cal
P}}-\bar{{\cal P}}{\cal P}-\bar{{\cal C}}^{\prime}{\cal
C}^{\prime}].
\end{eqnarray}
Performing integration over $\pi_1$, $p$, and $\bar{p}$ we land
onto the following generating functional.
\begin{eqnarray}
Z_{BR}&=&\int D\phi exp [i\int
d^2x[\pi_{\phi}\dot{\phi}+\pi_{\omega}\dot {\omega} +\pi_{0}(\dot
A_{0}-A_{1}^{\prime})+\frac{1}{2}\pi_{0}^{2}
+\frac{1}{2}{(\dot A_{1}-A_{0}^{\prime})}^2-2e^2A_{1}^2\nonumber\\
&-&\phi^{\prime2}+2e\phi^{\prime}(A_{0}-A_{1})- \alpha\omega'^2
-e(1+\alpha)A_0\omega' +2e\alpha(A_0+A_1)\omega' +\dot{\bar {\cal
C}\dot{\cal C}}-\bar{\cal C}^{\prime}{\cal C}^{\prime}]].
\end{eqnarray}
Note that the field $\pi_0$ is playing the role of
Nakanishi-Lautrup type auxiliary field. Let us call it as  $B$ for
convenience. The Lagrangian density now turns into
\begin{eqnarray}
{\cal L}_{eff}&=&\frac{1}{2}{(\dot
A_{1}-A_{0}^{\prime})}^2-2e^2A_{1}^2-\phi^{\prime2}
+\dot{\phi}\phi^{\prime} +2e\phi^{\prime}(A_{0}-A_{1})\nonumber \\
&+&\alpha(\dot{\omega}\omega' + \omega'^2) +
e(\alpha+1)(A_0\sigma'-A_1\dot{\omega}) -2e\alpha
(A_0+A_1)\omega+\frac{1}{2}{\cal B}^2 +{\cal B}\partial_{\mu}
A^{\mu} +\partial_{\mu}\bar{\cal C}\partial^{\mu}{\cal C}.
\label{EFFL}
\end{eqnarray}
The action corresponding to this theory is invariant under the
following transformations
\begin{eqnarray}
s_b\phi=-{\cal C} , s_b\omega={\cal C} ,
s_bA_0=\frac{1}{e}\dot{\cal C},s_b A_1=\frac{1}{e}{\cal C}^\prime,
s_b{\cal C}=0,s_b\bar{\cal C}=\frac{1}{e}{\cal B}, s_b{\cal B}=0.
\end{eqnarray}
Using the constraints of the theory we  can recast the BRST charge
in the following form
\begin{eqnarray}
\Omega
&=&\pi_{1}^{\prime}+e(\pi_{\phi}-\pi_{\omega})+e(\phi^{\prime}+\omega^{\prime}),
\nonumber\\
{\cal B}&=&\pi_0.
\end{eqnarray}
It is straightforward to define  define the anti-BRST charge for
this system:
\begin{equation}
{\cal Q}_{ab}=\frac{i}{e}[\Omega \bar{\cal C} -{\cal B}\dot{\bar
{\cal C}}], \label{ABQ}
\end{equation}
which generate the following ant-BRST transformations of the
fields describing the effective Lagrangian (\ref{EFFL})
\begin{eqnarray}
s_{ab}\phi=-\bar {\cal C}, s_{ab}\omega=\bar {\cal C},
s_{ab}A_0=\frac{1}{e}{\dot{\bar {\cal C}}},
s_{ab}A_1=\frac{1}{e}\bar {\cal C}^\prime, s_{ab}{\cal
C}=\frac{1}{e}{\cal B}, s_{ab}\bar{\cal C}=0, s_{ab}{\cal B}=0.
\end{eqnarray}
Let us now  calculate the equations of motion of the fields from
the Lagrangian (\ref{EFFL}) by using the Euler-Lagrange equation
in order to establish the important algebra between the nilpotent
charges.
\begin{equation}
\dot{\phi}^{\prime}-{\phi}^{''}+e(A_{0}^{\prime}-A_{1}^{\prime})=0,
\label{EB1}
\end{equation}
\begin{equation}
-\dot{\omega}^{\prime}-{\omega}^{''}+e(A_{0}^{\prime}+A_{1}^{\prime})=0,\label{EB2}
\end{equation}
\begin{equation}
\dot {\cal B}-\pi_1^{'}-2e(\phi^{'}+\omega^{'})=0,\label{EB3}
\end{equation}
\begin{equation}
-{\cal B}^{'}+\dot{\pi_1}+2e(\phi^{'}-\omega^{'})+4e^2
A_1=0,\label{EB4}
\end{equation}
\begin{equation}
{\cal B}+(\dot{A_0}-A_1^{'})=0,\label{EB5}
\end{equation}
\begin{equation}
\ddot{\bar{{\cal C}}}-\bar{{\cal C}}^{''}=0,\label{EB6}
\end{equation}
\begin{equation}
\ddot{{\cal C}}-{{\cal C}}^{''}=0.\label{EB7}
\end{equation}
The equations (\ref{EB1}), (\ref{EB2}), (\ref{EB3}), (\ref{EB4}),
(\ref{EB5}), (\ref{EB6}), and (\ref{EB7}) lead to the  following
useful relations
\begin{equation}
\dot{{\cal B}}={\Omega} , {{\cal B}}^{''}=\dot{\Omega}.
\end{equation}
It is straightforward to see that the BRST charge $Q_{b}$ and the
anti-BRST $Q_{ab}$ satisfy the following relations
\begin{eqnarray}
 \dot{{\cal Q}_b}=\dot{{\cal Q}_{ab}} =0, \lbrace {\cal Q}_b, {\cal Q}_{ab}
\rbrace=0, {\cal Q}_b^2=0, {\cal Q}_{ab}^2=0. \label{ALGB}
\end{eqnarray}
These algebras (\ref{ALGB}) guarantee the nilpotency of the BRST
and anti-BRST charges. It completes the   BRST and anti-BRST
property of the model under consideration.
\section{A discussion on the extended BRST symmetries of the model}
A careful look reveals that apart from BRST and anti-BRST symmetry
this model does have few other nilpotent symmetries. Let us now
proceed to explore that. We should mention here that although
these look nilpotent like the BRST symmetry, there are sharp
differences with the BRST symmetry and that plays a pivotal role
in the gauge symmetries of physically sensible theories. We are in
a position to examine the symmetries one by one with special
emphasis on the algebra of the charges corresponding to the
symmetries.
\subsection{Co-BRST and Anti-Co-BRST symmetry}
With this in view, we execute an investigation on the co-BRST and
anti-co-BRST symmetry properties of this model. The algebra
satisfied by the charges corresponding to these nilpotent
symmetries also has been studied exhaustively. It is beneficial to
mention at this stage that the total gauge fixing term remains
invariant under the co-BRST symmetry transformations along with
the invariance of the other terms involved in the theory. It is
known that the origin of the gauge fixing term remains encoded in
the co-exterior derivative $\delta=\pm *d*$ with $\delta^2 = 0$ of
differential geometry as the operation of $\delta$ on a one-form
produces the gauge-fixing term. The symbol $*$ indicates the Hodge
duality operation on the $2D$ spacetime manifold. The $\pm$ sign
refers to the dimensionality of the spacetime \cite{EGUCHI,
NISHI0, NISHI1}. Thus, the nilpotent co-BRST symmetry
transformations have their origin in the co-exterior derivative
$\delta$ of differential geometry. We find that co-BRST
transformations for the fields are
\begin{eqnarray}
s_d\phi=-\dot{\bar{{\cal C}}}, s_d\omega=\dot{\bar{{\cal C}}},
s_dA_0=\frac{1}{e}\bar{\cal C}^{''}, s_dA_1=\frac{1}{e}\dot{\bar
{{\cal C}}}^{\prime}, s_dc=\frac{1}{e}\Omega, s_d\bar{\cal C}=0,
s_d{\cal B}=0.
\end{eqnarray}
At this stage, it is straightforward  to write down the conserved
co-BRST charge of this theory which reads
\begin{equation}
{\cal Q}_d=\frac{i}{e}[\Omega {\dot{\bar {\cal
C}}}-\pi_0\bar{{\cal C}}^{''}].
\end{equation}
The action (\ref{EFFL}) is found to remain invariant under above
co-BRST transformation with a  little algebra.  The anti-co-BRST
charge is now written down as follows
\begin{equation}
{\cal Q}_{ad}=\frac{i}{e}[\Omega \dot{{\cal C}}-\pi_0 {{\cal
C}}^{''}].
\end{equation}
A careful look reveals that the charges ${\cal Q}_d$ and ${\cal
Q}_{ad}$ satisfy the following interesting relations.
\begin{eqnarray}
s_d {\cal Q}_d &=& -\lbrace{{\cal Q}_d,{\cal Q}_d}\rbrace=0, \nonumber\\
s_{ad} {\cal Q}_{ad}&=& -\lbrace {\cal Q}_{ad},{\cal Q}_{ad}\rbrace =0,\nonumber\\
s_{d} {\cal Q}_{ad}&=& -\lbrace {\cal Q}_{ad},{\cal Q}_{d}\rbrace =0,\nonumber\\
s_{ad} {\cal Q}_{d}&=& -\lbrace {\cal Q}_{d},{\cal Q}_{ad}\rbrace =0.\nonumber\\
\end{eqnarray}
\subsection{Bosonic symmetry}
We have already discussed the BRST, anit-BRST, co-BRST, and
anti-co-BRST symmetry. Besides, it is found that this theory has
one more symmetry. This symmetry is indeed constituted with the
aforesaid discussed BRST, anti-BRST, co-BRST, and anti-co-BRST
symmetry. It is found that the following relations are satisfied.
\begin{eqnarray}
&&\lbrace s_d, s_{ad}\rbrace=0, \nonumber\\
&&\lbrace s_b, s_{ab}\rbrace=0, \nonumber\\
&&\lbrace s_b, s_{ad}\rbrace=0, \nonumber\\
&&\lbrace s_d, s_{ab}\rbrace=0 \nonumber\\
&&\lbrace s_b, s_{d}\rbrace=s_w, \nonumber\\
&&\lbrace s_{ab}, s_{ad}\rbrace=s_{\bar w}.
\end{eqnarray}
Here $w$ corresponds to the bosonic symmetry. The field variables
have the following transformations under the bosonic symmetry
transformation:
\begin{eqnarray}
s_w\phi&=&-\frac{i}{e}(\dot{{\cal B}}+\Omega),\nonumber\\
s_w\omega&=&\frac{i}{e}(\dot{{\cal B}}+\Omega),\nonumber\\
s_w{A_0}&=&\frac{i}{e}({{\cal B}}^{''}+\dot{\Omega)},\nonumber\\
s_w{A_1}&=&\frac{i}{e}(\dot{{\cal B}^{'}}+{\Omega}^{\prime}) ,\nonumber\\
s_w {\cal C}&=&0, \nonumber\\
s_w{\bar {\cal C}}&=&0, \nonumber\\
s_w {\cal B}&=&0.
\end{eqnarray}
However, the symmetry corresponding to  $s_{\bar w}$ is not
independent  because we find that the operation of  $s_w$ and
$s_{\bar w}$  have the following linear algebraic relations
between themselves:
\begin{eqnarray}
s_w+s_{\bar w}=0, ie, \lbrace s_b, s_{d}\rbrace=s_w=-\lbrace
s_{ab}, s_{ad}\rbrace
\end{eqnarray}
This symmetry of course has a  conserved charge. The conserved
charge corresponding to this bosonic symmetry reads
\begin{equation}
{\cal Q}_w=\frac{i}{e^{2}}(\Omega^2 - {\cal B}{\cal B}^{''}).
\end{equation}
With the help of the equations of motion we land onto the
following useful relations:
\begin{eqnarray}
{\dot{\cal Q}}_w = \frac{d{\cal
Q}_w}{dt}=-\frac{i}{e^{2}}[\dot{{\cal B}}{\cal B}^{''}+{\cal
B}\dot{{\cal B}^{''}}] =\frac{i}{e^{2}}[\dot{{\cal B}}{\cal
B}^{''} -{\cal B}\dot{{\cal B}^{''}}]=0,
\end{eqnarray}
since $\Omega\approx 0$, and consequently $\dot{\Omega}\approx 0$.
Therefore it appears the bosonic charge ${\cal Q}_w$ is a constant
of motion of the theory. We have already seen that the theory is
endowed with the BRST, anti-BRST, co-BRST, anti-co-BRST symmetry
along with a bosonic symmetry. Apart from the presence of these
symmetries, the theory has some extra symmetries. We now turn to
observe that.
\subsection{Ghost and discrete symmetry}
We know that ghost and anti-ghost fields are designated by a
specific number called ghost numbers. For the ghost field, the
ghost number is $1$ and the corresponding number for the
anti-ghost field is $-1$. The matter, anti-matter, and gauge field
of course have zero ghost numbers. The above fact provides a scale
transformation that keeps the effective action of the theory
invariant. We introduce the following scale transformation of the
ghost field
\begin{eqnarray}
\phi\rightarrow\phi,\omega\rightarrow\omega,A_0 \rightarrow
A_0,A_1 \rightarrow A_1,{\cal B} \rightarrow {\cal B},{\cal C}
\rightarrow {e^{\lambda}{\cal C}},\bar{{\cal C}}\rightarrow
{e^{-\lambda}}\bar{{\cal C}},
\end{eqnarray}
where $\lambda$ is a global scale parameter and we find that the
effective action of the theory remains invariant under these
transformations. The above transformations in the infinitesimal
limit take the following form
\begin{eqnarray}
s_g\phi=0,s_g\omega=0,s_g A_0=0, s_g A_1=0, s_g B=0, s_g {\cal
C}={\cal C}, s_g \bar{{\cal C}}=-\bar{{\cal C}}.
\end{eqnarray}
According to Noether's theorem, this symmetry must have a
conserved charge and that conserved charge for this ghost symmetry
is given by
\begin{equation}
{\cal Q}_g=i[\dot{\bar{{\cal C}}}{\cal C}+\dot{{\cal C}}\bar{{\cal
C}}].
\end{equation}
The ghost sector is found to  respect a discrete symmetry in
addition to the above continuous symmetry transformation:
\begin{eqnarray}
 {\cal C} \rightarrow \pm i
\bar{{\cal C}}, ~~~~~\bar{C}\rightarrow \pm i{\cal C} .
\end{eqnarray}
This ends the discussion of the symmetry properties of this model.
It is more interesting to investigate the geometrical cohomology
corresponding to the symmetry of this model under consideration.
\section{Cohomological aspect of the theory}
In differential geometry, de Rham cohomological operators are
known to obey the following important algebra
\begin{equation}
d^2=\delta^2=0, \Delta=(d+\delta)^2= d\delta + \delta d
=[d,\delta]_{-},
\end{equation}
\begin{equation}
[\Delta,\delta]=0, [\Delta, d]=0.
\end{equation}
Here $d$ and $\delta$ are respectively known as exterior and co
exterior operator and $\Delta$ is known Laplace-Beltrami operator.
Let us now look carefully at the algebra of the conserved charges
corresponding to all these symmetries which the theory is
possessing:
\begin{eqnarray}
&&{\cal Q}_b ^2 =0, {\cal Q}_{ab} ^2=0, Q_d ^2=0, {\cal Q}_{ad}^2=0\nonumber \\
&& \lbrace Q_b,Q_{ab}\rbrace=\lbrace Q_d, Q_{ad}\rbrace=\lbrace
{\cal Q}_b,{\cal Q}_{ad}\rbrace=\lbrace {\cal Q}_d, {\cal
Q}_{ab}\rbrace=0, \nonumber \\
&&[{\cal Q}_g, {\cal Q}_b]= {\cal Q}_b, [{\cal Q}_g, {\cal
Q}_{ab}]=-{\cal Q}_{ab}, [{\cal Q}_g, {\cal Q}_d]=-{\cal Q}_d,
[{\cal Q}_g, {\cal Q}_{ad}]={\cal Q}_{ad}, \nonumber \\
&&{\cal Q}_w  = -\lbrace {\cal Q}_b,{\cal Q}_d\rbrace=\lbrace
{\cal Q}_{ad}, {\cal Q}_{ab}\rbrace, [{\cal Q}_w,{\cal
Q}_{\alpha}]=0 \label{QAL} .
\end{eqnarray}
where ${\cal Q}_{\alpha}\equiv ({\cal Q}_b,{\cal Q}_{ab},{\cal
Q}_d,{\cal Q}_{ad},{\cal Q}_g)$. The above relations (\ref{QAL})
transpires that ${\cal Q}_w$ is the Casimir operator of the whole
algebra.

We know that in differential geometry the role of an exterior
derivative is to raise the degree of a form by one, i.e., $df_n=
f_{n+1}$ whereas the role of the co-exterior derivative is the
reverse, it lowers the degree of a form by one, i.e., $\delta f_n=
f_{n-1}$. Here $f_n$ represents a form of degree $n$. Under the
operation of $\Delta$ however, the degree of a form remains
unaltered. Let us now define a state $\vert\eta \rangle$ with
ghost number $\kappa$ in the Hilbert space of states of this BRST
invariant theory as
\begin{equation}
i{\cal Q}_g {\vert \eta \rangle}_\kappa=\kappa{\vert \eta
\rangle}_\kappa.
\end{equation}
It is  straightforward to  verify the relations
\begin{eqnarray}
i{\cal Q}_g {\cal Q}_b {\vert \eta
\rangle}_\kappa&=&(\kappa+1){\cal Q}_b, {\vert \eta
\rangle}_\kappa\nonumber \\
i{\cal Q}_g {\cal Q}_{ad} {\vert \eta \rangle}_\kappa&=&(\kappa
+1)Q_{ad} {\vert \eta
\rangle}_\kappa,\nonumber \\
i{\cal Q}_g {\cal Q}_d {\vert \eta \rangle}_\kappa&=&(\kappa-1)
{\cal Q}_d {\vert \eta \rangle}_\kappa,
\nonumber \\
i{\cal Q}_g {\cal Q}_{ab} {\vert \eta
\rangle}_\kappa&=&(\kappa-1){\cal Q}_{ab} {\vert \eta
\rangle}_\kappa,\nonumber \\
i{\cal Q}_g {\cal Q}_w {\vert \eta \rangle}_\kappa&=&\kappa {\cal
Q}_w {\vert \eta \rangle}_\kappa. \label{DR}
\end{eqnarray}
A careful look into  the equation (\ref{DR}) transpires the
following analogy:
\begin{equation}
({\cal Q}_b,{\cal Q}_{ad})\rightarrow d,({\cal Q}_d,{\cal
Q}_{ab})\rightarrow \delta, {\cal Q}_w\rightarrow \Delta.
\end{equation}
Note that $({\cal Q}_b,{\cal Q}_{ad})$ raise the ghost number of
the state by one whereas $({\cal Q}_d,{\cal Q}_{ab})$ lower the
ghost number by one, and  ${\cal Q}_w$ keeps the ghost number
unchanged. So $({\cal Q}_b,{\cal Q}_{ad}), ({\cal Q}_d,{\cal
Q}_{ab}),{\cal Q}_w$ resemble the algebra of $ d, \delta, \Delta$.
A closer look reveals that the analogy with the Hodge-de Rham
decomposition theorem enables us to express any arbitrary state
${\vert \eta \rangle}_n$ in terms of the sets $({\cal Q}_b,{\cal
Q}_d,{\cal Q}_w)$ and $({\cal Q}_{ad},{\cal Q}_{ab},{\cal Q}_w)$
as
\begin{equation}
{\vert \eta\rangle}_\kappa= {\vert \sigma \rangle}_\kappa +{\cal
Q}_b {\vert \chi \rangle}_{\kappa-1}+ {\cal Q}_d {\vert\rho
\rangle}_{\kappa+1},
\end{equation}
\begin{equation}
{\vert \eta \rangle}_\kappa= {\vert \sigma \rangle}_\kappa +{\cal
Q}_{ad} {\vert \chi \rangle}_{\kappa-1}+ {\cal Q}_{ab} {\vert \rho
\rangle}_{\kappa+1},
\end{equation}
where the most symmetric state is the harmonic state ${\vert
\sigma \rangle}_\kappa$, that satisfies the following equations
\begin{equation}
{\cal Q}_w{\vert \sigma \rangle}_\kappa=0, {\cal Q}_b{\vert \sigma
\rangle}_\kappa=0,{\cal Q}_d{\vert \sigma \rangle}_\kappa=0,{\cal
Q}_{ab}{\vert \sigma \rangle}_\kappa=0, {\cal Q}_{ad}{\vert \sigma
\rangle}_\kappa=0.
\end{equation}
The charges ${\cal Q}_b,{\cal Q}_{ab},{\cal Q}_d$, and ${\cal
Q}_{ad}$ corresponding to the symmetries of  a physically sensible
theory must maintain the conditions
\begin{eqnarray}
{\cal Q}_b \vert Phy \rangle=0,~~ {\cal Q}_{ab} \vert Phy
\rangle=0,
\nonumber\\
{\cal Q}_d \vert Phy \rangle=0,~~ {\cal Q}_{ad} \vert Phy
\rangle=0,
\end{eqnarray}
which lead to the following two conditions on the first-class
constraints that generate the gauge as well as BRST symmetry of
the theory:
\begin{equation}
\pi_0 \vert Phy \rangle=0,ie, ~~{\cal B} \vert Phy \rangle=0
\end{equation}
\begin{equation}
[\pi_{1}^{\prime}+e(\pi_{\phi}-\pi_{\omega})+e(\phi^{\prime}+\omega^{\prime})]\vert
Phy \rangle=0,ie, ~~ \Omega \vert Phy \rangle= 0.
\end{equation}
\section{Summary and discussion}
We have taken into consideration a gauged Lagrangian with a Siegel
type chiral boson with a parameter-involved non-covariant masslike
term for the gauge field. The masslike term that was selected in
\cite{BEL} resulted in a gauged theory of Floreanini-Jackiw type
chiral boson which may be derived from the Chiral Schwinger model
with the Jackiw-Rajaraman type of the electromagnetic anomaly as
became proven with the incredible concept of imposition of chiral
constraint in the article of Harada \cite{KH}. To derive the
gauged version of Floreanini-Jackiw type of chiral boson a
parameter involved masslike term is brought right here in Eqn. (1)
with the anticipation that we receive the same gauged model of
chiral boson \cite{ARC} obtained from the chiral Schwinger model
with the one-parameter class of improved Faddeevian anomaly
\cite{WOT}.

In the article \cite{PM}, the author showed that the chiral
Schwinger model remains physically sensible in all respect with a
parameter-unfastened masslike term where the nature of anomaly
belonged to the Faddeevian class and a one parameter-class of
improved Faddeevian anomaly was generated in \cite{WOT}. We have
made here an equivalence between the gauge-invariant and gauge
non-invariant version of this gauged Floreanini-Jackiw type chiral
boson with a parameter-involved masslike term using the ingenious
formalism developed in \cite{FALCK}. The role of gauge fixing is
found very crucial here because an arbitrary but legitimist gauge
fixing may lead to other effective theories which may fail to
establish the equivalence. Using the formalism developed in
\cite{FIK} we write down the BRST invariant effective action
without going through the formal BFV development of BRST
quantization as it was followed in the article \cite{RPM0, RPM4,
HODGE5} to study the BRST symmetry and the corresponding BRST
cohomology. In the article \cite{RPM4} however, BRST cohomology
was not attempted. We study the extended version of the BRST
algebra. We also notice that the nilpotent anti-BRST symmetry
transformations always satisfy the absolute anti-commutativity.
The nilpotency signalizes the fermionic nature of the anti-BRST
symmetry transformations and the absolute anti-commutativity
encodes the linear independence of these transformations. The
gauge fixing term, owing to its origin in the co-exterior
derivative remains invariant under co-BRST symmetry
transformations. Thus, the co-exterior derivative can be realized
in terms of the co-BRST symmetry of the theory. The
anti-commutator of BRST and co-BRST transformations produce a
bosonic symmetry which is an analog of the Laplacian operator. It
is also found that the ghost terms of the theory remain invariant
under the bosonic symmetry transformations. Finally, we have shown
that, at the algebraic level, the aforesaid symmetry
transformations resemble the identical algebra as it is found to
be satisfied by the de Rham cohomological operators of
differential geometry.

BRST quantization of the parameter-unfastened Faddeevian anomaly
in a formal way using BFV was done earlier in \cite{AR1} where we
did not study the cohomological aspects. In this article, our main
emphasis is to study the cohomological aspects that precisely
require an extended version of the BRST algebra i.e., the algebra
between the different transformation generators. To be precise,
these are the BRST, anti-BRST, co-BRST, anti-co-BRST, and the
conserved charge corresponding to the bosonic symmetry. We have
been succeeded to establish that the model with
parameter-dependent Faddeevian anomaly also belongs to the Hodge
class.

One of the highlights of our present investigation is the
observation that the Lagrangian density under consideration
provides a model for the Hodge theory because the continuous
symmetry operators of the specific Lagrangian density and the
corresponding charges obey an algebra that is reminiscent of the
algebra obeyed by the de Rham cohomological operators of
differential geometry. In other words, the continuous symmetry
operators provide the physical realizations of the cohomological
operators of differential geometry. This happens because of the
fact that the Lagrangian density respects five perfect symmetries.
This is precisely the reason that four of the above-mentioned five
symmetries of the theory obey an exact algebra that a reminiscent
of the algebra obeyed by the de Rham cohomological operators of
the differential geometry.

A cautious look exhibits that the version developed in \cite{KH}
and \cite{PMS} appear with quite different mathematical and
structural forms. However these two \cite{KH} and \cite{PMS} are
essentially originated from the fermionic chiral Schwinger model
which was described in \cite{JR}. A precise bosonization of this
fermionic model within the $(1+1)$ dimensional has been found
possible. A one-loop correction enters inside that via
regularization during the course of the bosonization process. This
regularization may be executed in unique methods and that ends up
with unique counter terms. Parameter-unfastened Faddeevian
regularization came in the literature due to Mitra \cite{PM}. The
generalization with the one-parameter involved improved Faddeevian
regularization was developed in \cite{WOT}. It has been observed
that each of the models may be expressed in terms of chiral boson
\cite{PMS, ARC}, like Jackiw-Rajaraman's version of the chiral
Schwinger model which was offered by Harada \cite{KH}. So it
exhibits that those versions are the outcome of the use of various
regularizations during the course of bosonization of the fermionic
model proposed in \cite{JR}. Therefore, it has been found that
unique regularization ends in the bosonized version that has an
appearance quite unique in a structural sense. Most of these
bosonized versions can be expressed in terms of chiral bosons and
they have identical symmetry structure at the quantum level and
they all belong to the magnificent Hodge class. In the article
\cite{ASE}, it has been visible that for the bosonized model of
the chiral Schwinger with different regularizations (the standard
one and the parameter-unfastened Faddeevian) and also for the
Schwinger model \cite{HODGE5} the symmetry at the quantum level
stays unchanged and it does no longer rely on the selection of
regularization and certainly, the Hodge algebra becomes satisfied
with the aid of the extended BRST symmetries. Here we observe that
the bosonized model of the chiral Schwinger model with the
one-parameter class of improved Faddeevian regularization
additionally, respect the identical symmetry and belongs to the
same magnificat Hodge class. So via case studies, it has been has
been established that the symmetry of the chiral Schwinger model
stays identical at the quantum level regardless of the character
of anomaly. Of course, extra distinct research is needed to
establish it in general.

\end{document}